\documentclass[prd,amsfonts,noshowpacs,nofootinbib]{revtex4}
\usepackage{amsmath}
\usepackage{multirow}
\usepackage{chngpage}
\usepackage{graphicx,subfigure,epsfig}

\def\be{\begin{equation}}
\def\ee{\end{equation}}
\def\bea{\begin{eqnarray}}
\def\eea{\end{eqnarray}}
\newcommand{\rs}{\triangle_{\cal{R}}^{4}}
\renewcommand{\H}{{\cal H}}
\renewcommand{\k}{{\bf k}}
\newcommand{\x}{{\bf x}}
\newcommand{\rsn}{\triangle_{\cal{R}}^{2}}

\begin{document}
\title{\bf{Constraints on primordial density perturbations from induced gravitational
waves}}
\author{Hooshyar Assadullahi and David Wands}
\affiliation{Institute of Cosmology and Gravitation, University of
Portsmouth, Dennis Sciama Building, Burnaby Road, Portsmouth PO1
3FX, United Kingdom}

\begin{abstract}
We consider the stochastic background of gravitational waves
produced during the radiation-dominated hot big bang as a constraint
on the primordial density perturbation on comoving length scales
much smaller than those directly probed by the cosmic microwave
background or large-scale structure. We place weak upper bounds on
the primordial density perturbation from current data. Future
detectors such as BBO and DECIGO will place much stronger
constraints on the primordial density perturbation on small scales.
\end{abstract}

\maketitle

\section{Introduction}

Recent cosmic microwave background (CMB) experiments
\cite{Komatsu:2008hk,Hinshaw:2008kr} and high-redshift galaxy
surveys \cite{Reid:2009xm} are able to probe the primordial density
perturbation on cosmological scales, $\sim10-1000$~Mpc. This yields
a precise measure of the primordial density contrast,
$\rsn(k_{CMB})=2.5\times10^{-9}$ for $k_{CMB}=0.002$~Mpc$^{-1}$
\cite{Komatsu:2008hk}.
For much smaller scales, free-streaming of relativistic particles in
the very early universe and Silk damping due to photon diffusion
erases the primordial density perturbation on comoving scales much
less than 10~Mpc~\cite{Peacock:1999ye}. The present distribution of
matter on smaller scales is the result of the subsequent non-linear
evolution of the matter density and we are unable to directly relate
observational data to the primordial distribution of matter.
Therefore for modes which are much smaller than 10~Mpc the
primordial density perturbation remains undetermined.

One might assume that the primordial density perturbation on small scales has
the same value as on larger scales. According to WMAP 5-year data
\cite{Komatsu:2008hk,Hinshaw:2008kr} the spectral index on CMB
scales is smaller than one, $n_{s}=0.96^{+0.014}_{-0.013}$),
implying that the density contrast is slightly smaller on smaller
scales, but it is a huge extrapolation to assume that this is the
actual scale-dependence all the way from CMB scales down to comoving
scales such as $k^{-1}<10$~pc, for instance, which are smaller than
the comoving Hubble scale at the epoch of big bang nucleosynthesis,
$T\sim$~MeV. The only probes of such small scales at early times are
expected to be gravitational relics.

%Even if we can not directly determine the primordial density
%perturbation on small scales, we can place bounds upon it.

One process which allows us to put an upper bound on density
perturbations is the formation of primordial black holes (PBHs)
\cite{Carr:1974nx}. PBHs are produced when density fluctuations with
a large amplitude (${\delta\rho}/{\rho}\approx 0.01 -0.1$) enter the
horizon. Their typical mass is given by the horizon mass when they
were produced
\cite{Green:1997sz,Lyth:2005ze,Carr:2005bd,Kohri:2007qn}.
\begin{equation}
\label{PBHM} M_{PBHs}\approx 10^{15}\left(\frac{t}{10^{-23}\rm
s}\right) \rm g
\end{equation}
which can be related to the temperature if they formed in the
radiation dominated era
\begin{equation}
\label{PBHMT} M_{PBHs}\approx 10^{38}\left(\frac{\rm
MeV}{T}\right)^2 \rm g
\end{equation}
The fact that PBHs have not been observed to date limits the initial
mass fraction going into black holes, which can be used to constrain
the primordial density perturbation on the corresponding scales
\cite{Green:1997sz}. For instance PBHs radiate Hawking radiation and
thus for $M_{PBHs}<10^{15}$g evaporations are limited because of
constraints from the standard big bang nucleosynthesis (BBN), which
is sensitive to the baryon-photon ratio at the time, and also
because of bounds on the expected gamma-ray background.
%
% This places an upper bound on the density perturbation on the
% corresponding scales but it does not make a tight constraint on
% primordial density perturbation ($\frac{\delta\rho}{\rho}\approx
% 0.01 -0.1$ on the relevant scale \cite{Saito:2008jc}).
%
For heavier PBHs ($M_{PBHs}>10^{15}$g), similar constraints come
with the fact that the present-day density of PBHs cannot exceed the
upper limit on the cold dark matter density \cite{Zaballa:2006kh}.

%because of the energy density of PBHs in comparison with energy
%density of universe, so there is a limit for the number of PBHs in
%the universe $n<\frac{3H_{0}^{2}}{8\pi G}M_{PBHs}$

Gravitational waves also give us a window onto density perturbations
in the very early universe. They can be generated by violent events
in the early universe such as  bubble collisions
\cite{Kamionkowski:1993fg}, cosmic strings \cite{Allen:1991bk},
preheating after inflation
\cite{Khlebnikov:1997di,Easther:2006vd,Easther:2007vj,Dufaux:2008dn},
or parametric decay of supersymmetric condensates
\cite{Dufaux:2009wn,Kusenko:2009cv}. All of these mechanisms are
highly model-dependent. Even gravitational waves from produced from
vacuum fluctuations of the metric during inflation
\cite{Starobinsky:1979ty,Turner:1996ck,Maggiore:1999vm}, though a generic prediction of
inflation, have an amplitude which is dependent upon the energy
scale of inflation.

In this paper we consider the bounds placed on the the primordial
density perturbation from the generation of a stochastic background
of gravitational waves in the very early universe.
In the standard radiation-dominated hot big bang, first-order
density perturbations inevitably generate gravitational waves at
second (and higher) order
% v3 extra refs added
\cite{Tomita:1967,Matarrese:1993zf,Matarrese:1996pp,Matarrese:1997ay,Noh:2004bc,Carbone:2004iv,Nakamura:2004rm,Ananda:2006af,Baumann:2007zm,
Mangilli:2008bw,Malik:2008im}. Like the initial mass fraction
of PBHs, the power spectrum of induced gravitational waves is
determined by the primordial density perturbation and thus can place
bounds on the amplitude of scalar perturbations. Indeed Saito and
Yokoyama \cite{Saito:2008jc} have recently pointed out that current
gravitational wave bounds are sufficient to rule out PBHs as
possible candidates for intermediate mass black holes.

If we could determine the amplitude and frequency of these induced
gravitational waves then we would be able to determine the
primordial density perturbation when the corresponding scales
crossed the Hubble scale during the early hot big bang.
Throughout this paper we will use the following formula to relate
the frequency of gravitational waves at the present time to the
temperature at Hubble-crossing in the early radiation-dominated
era~\cite{Assadullahi:2009nf}:
\begin{equation}
\label{fT} \nu = \frac{c}{\lambda_0} \approx 1.2 \times 10^{-8}\,
g_*^{1/6} \left( \frac{T}{{\rm GeV}} \right)\, {\rm Hz} \,,
\end{equation}
where $g_*$ is the effective number of degrees of freedom.
We are able to place upper bounds on the primordial density
perturbation from BBN
and cosmic microwave background (CMB) constraints and as well as current LIGO, VIRGO
and pulsar timing data. If we do not detect gravitational waves
with future pulsar timing \cite{Jenet:2005pv,Jenet:2006sv} and
future detectors like Advanced LIGO \cite{Abbott:2006zx}, Advanced
VIRGO \cite{Acernese:2008zzf}, LISA \cite{Hughes:2007xm}, BBO
\cite{Corbin:2005ny} and DECIGO \cite{Kawamura:2008zz}, the upper
bounds on the primordial density perturbation will become
significantly tighter.

% v3
We emphasize that our bounds come from adopting the standard, minimal cosmological model of adiabatic density perturbations, in their growing mode, in a radiation-dominated early universe from ultra-high energies ($\sim 10^{16}$~GeV) until matter-domination when $T<$~eV. The quantitative constraints will be altered if one adopts non-standard cosmological evolution \cite{Seto:2003kc,Boyle:2005se} such as an early matter-era (e.g., temporary domination of the energy density by massive, non-relativistic particles) or a stiff-fluid-dominated era (e.g., domination by the kinetic energy of a coherent, fast-rolling scalar field).
On general grounds one expects an early era dominated by fluid "softer" than radiation, $P/\rho<1/3$, to dilute the fractional density of gravitational waves whose wavelength is smaller than the comoving Hubble scale (and thus behave like radiation), while the fractional density of sub-horizon gravitational waves grows relative to matter which is stiffer than radiation \cite{Giovannini:2008zg}. On the other hand the evolution of density perturbations which give rise to gravitational waves is also altered, see for example Ref.~\cite{Assadullahi:2009nf}. Non-adiabatic modes in a multi-component system can lead to large-scale adiabatic density perturbations by the time of last-scattering but arise from initial isocurvature perturbations, so are not necessarily constrained by our analysis. One such example is the curvaton scenario, where the gravitational waves may be enhanced with respect to the adiabatic case if the curvaton is subdominant when it decays~\cite{Bartolo:2007vp}. Such models need to be considered on a case by case basis.

This paper is organized as follows: in section~2 we introduce the
basic equations used to determine the induced gravitational wave
background and define the effective energy density of second-order
gravitational waves. In section~3 we quantify the constraints placed
on the primordial density perturbation by a variety of experiments.
We present our conclusions in section~4.

\section{Second-order gravitational waves}

In this section we will briefly review the generation of induced
gravitational waves. Details of the calculations have been described
previously \cite{Ananda:2006af,Baumann:2007zm}.

The perturbed metric in the longitudinal gauge is
\begin{equation}
 \label{metric}
  ds^{2}
   =
 a^{2}(\eta)[-(1+2\Phi)d\eta^{2}+[(1-2\Psi)\delta_{ij}+2F_{(i,j)}+h_{ij}]dx^{i}dx^{j}]
\end{equation}
where $\Phi$ and $\Psi$ are scalar metric perturbation, $F_{i}$ is a
transverse vector perturbation and $h_{ij}$ is a transverse and
trace-free tensor perturbation. The scalar metric perturbations,
$\Phi$ and $\Psi$, are supported by density perturbations, and in
the absence of anisotropic stress we require $\Phi=\Psi$
\cite{Mukhanov:1990me}. We will find it convenient to use the
Fourier transform
\begin{equation}
 \label{fourier}
 \Phi(\x) = \frac{1}{(2\pi)^{\frac{3}{2}}} \int d^3\k\, \Phi_{\k}
\, e^{i\k.\x} \,,
\end{equation}
where, for an isotropic distribution, the power spectrum is given by
\begin{equation}
 \label{scalar power}
 \langle {\Phi}_{\k} {\Phi}_{\k'} \rangle
 = \frac{2\pi^{2}}{k^{3}} \delta^3(\k+\k') \, {\cal {P}}(k)
\end{equation}
On large scales (much larger than the Hubble scale) the power
spectrum of the primordial scalar perturbation is commonly
approximated by a power law
\begin{equation}
\label{ns-later} {\cal{P}}(k) =
% v5 changed \frac{9}{25}
\frac49 \triangle_{\cal{R}}^{2}
 \left( \frac{k}{k_*} \right)^{n_{s}-1}
\end{equation}
where the numerical factor
% v5 $9/25$
$4/9$ comes from the relation between
scalar metric perturbation in the longitudinal and comoving gauges
on large scales in a
% v5 matter-dominated
radiation-dominated era \cite{Mukhanov:1990me}.

In the second-order perturbed Einstein field equations we see the
effect of first-order perturbations as a source term ($S_{ij}$) for
second-order tensor perturbations. After putting all first-order
perturbation terms to the right-hand side of the Einstein field
equation, we have
\begin{equation}
\label{h''1}
 h''_{ij}+2{\cal H}  h'_{ij}+k^{2}h_{ij}=S_{ij}^{TT}
 \,,
\end{equation}
where $S_{ij}^{TT}$ indicates the transverse-tracefree part of the
source term.
If we neglect first order tensor and vector perturbations in
comparison with first order density perturbations,  the right hand
side of this equation is the transverse and trace-free part
quadratic in first-order scalar perturbations. This  behaves like a
source term for induced gravitational waves
\cite{Ananda:2006af,Baumann:2007zm}
\begin{eqnarray}
\label{s1} \nonumber S_{ij} &=&
 2\Phi\partial_{i}\partial_{j}\Phi - 2\Psi\partial_{i}\partial_{j}\Phi
 + 4\Psi\partial_{i}\partial_{j}\Psi + \partial_{i}\Phi\partial_{j}\Phi
 - \partial^{i}\Phi\partial_{j}\Psi - \partial^{i}\Psi\partial_{j}\Phi
 + 3\partial^{i}\Psi\partial_{j}\Psi\\
\nonumber&&
 - \frac{4}{3(1+w){\cal H}^{2}}\partial_{i}(\Psi'+{\cal
H}\Phi)\partial_{j}(\Psi'+{\cal H}\Phi)\\
&&
 -\frac{2c_{s}^{2}}{3w{\cal H }}\,[3{\cal H}({\cal
H}\Phi-\Psi')+\nabla^{2}\Psi]\,\partial_{i}\partial_{j}(\Phi-\Psi)
%\\
\end{eqnarray}
where $w=P/\rho$ is the equation of state and $c_s^2=P'/\rho'$ is
the adiabatic sound speed.

These equations are written in the real space but in order to derive
the power spectrum of gravitational waves we need to transform to
Fourier space \cite{Ananda:2006af}
\begin{equation}
\label{tensore fourier}
 h_{ij}(x,\eta) =
 \int\frac{d^{3}\k} {(2\pi)^{\frac{3}{2}}} e^{i\k.\x}[h_{\k}(\eta)e_{ij}(\k)+\bar{h}_{\k}\bar{e}_{ij}(\k)]
 \,,
\end{equation}
where $e^{ij}(\k)$ and $\bar{e}_{ij}(\k)$ are the polarization
tensors. The two polarization tensors $e_{ij}(k)$ and
$\bar{e}_{ij}(k)$ can be given in terms of the orthonormal basis
\begin{eqnarray}
\label{polarization} \nonumber e_{ij}(\k) &=&
 \frac{1}{\sqrt{2}}[e_{i}(\k)e_{j}(\k)-\bar{e}_{i}(\k)\bar{e}_j(\k)]\\
 \bar{e}_{ij}(\k) &=& \frac{1}{\sqrt{2}}[e_{i}(\k)\bar{e}_j(\k)+\bar{e}_{i}(\k)e_{j}(\k)]
 \,,
\end{eqnarray}
where ${\bf e}$ and ${\bf \bar{e}}$ are unit vectors orthogonal to
one another and $\k$:
\begin{equation}
\label{ek} e_{i} k^{i}=\bar{e}_{i} k^{i}=e_{i}\bar{e}^{i}=0
\end{equation}

The gravitational waves have a power spectrum in Fourier space
\begin{equation}
 \label{tensor power}
 \langle h_{\k}(\eta)h_{\k'}(\eta) \rangle
 = \frac12 \frac{2\pi^{2}}{k^{3}}\delta^3(\k+\k'){\cal {P}}_{h}(k,\eta)
 \,,
\end{equation}
The effective density of a stochastic background of gravitational
waves, on scales much smaller than the Hubble scale, is given by
\cite{Maggiore:1999vm}
\begin{equation}
\label{maggiore} \rho_{GW} = \frac{1}{32\pi G} \langle
\dot{h}_{ij} \dot{h}^{ij} \rangle = \frac{k^2}{32\pi G a^2} \int
d(\ln k)\ {\cal P}_h(k,\eta) \,.
\end{equation}
The fraction of the critical energy density in gravitational waves
per logarithmic range of wavenumber $k$ in the radiation era is thus
\begin{equation}
\label{omegap}
 \Omega_{GW}(k,\eta) = \frac{1}{12} \left( \frac{k}{\H} \right)^2 {\cal P}_h(k,\eta)
  \,.
\end{equation}
After the radiation-dominated era, the density of gravitational
waves on sub-Hubble scales then redshifts exactly as any
non-interacting relativistic particle species and in the present day
we have
\begin{equation}
 \label{OmegaGW0}
 \Omega_{GW,0}(k)=\frac {\Omega_{\gamma,0}}{12} \left( \frac{k}{\H} \right)^2 {\cal P}_h(k,\eta)
  \,,
\end{equation}
where we
neglect additional numerical factors due to the detailed thermal
history, such as the heating of photons by the annihilation of other
relativistic particle species \cite{Boyle:2005se,Nakayama:2008wy}.
% v4
The present density of photons is $\Omega_{\gamma,0}\simeq
%2.48\times10^{-5}h^{-2}
4.8\times10^{-5}$ where here, and throughout this paper, we take $H_{0}\simeq72$~km~s$^{-1}$Mpc$^{-1}$ for the present value of the Hubble rate.

% The existence of density perturbations, generate induced
% gravitational waves. Therefore they can be generated in different
% eras in the early universe. In the  next section we will
% investigate the generation of induced gravitational waves during
% the radiation era.

\section{Constraints on primordial density perturbations}

In the standard cosmological scenario, second-order gravitational
waves are generated during the radiation-dominated era after
inflation.
% v5
Non-linear interactions can in principle lead to density perturbations integrated over a range of scales contributing to the gravitational wave amplitude on a given wavenumber, $k$, but in practice the second-order gravitational waves are primarily produced when first-order density perturbations on the
similar on same scale, $\sim k$, come inside the Hubble scale during the radiation era \cite{Ananda:2006af}.

%Baumann et al \cite{Baumann:2007zm} continued this work, and found
%gravitational waves in the matter dominated universe after the
%radiation era, however in this section we only consider the
%production of induced gravitational waves during the radiation
%era. Although \cite{Baumann:2007zm}, denotes that on the very
%large scale during the late matter dominated era, we have a
%prominent increase in the amplitude of gravitational waves, but
%this effect happens for the modes which are just inside the
%horizon at the present time, clearly these modes are to big to be
%observed with the direct detectors so we are unable to make any
%constraints on the density perturbation with them.
%
% v5
Assuming a power-law spectrum for the primordial density perturbation, Eq.~(\ref{ns-later}),
the energy density of second-order gravitational waves, relative to
the critical density at the present time, which were produced during
the radiation-dominated era ($\nu>10^{-15}$ Hz), can be written as
\begin{equation}
\label{hr1}
%\Omega_{gw,0}(k)=\Omega_{\gamma,0}\times\frac{8}{3}\times\frac{216^{2}}{\pi^{3}}\times
%F\times\rs(k)
\Omega_{gw,0}(k)=F_{\rm rad}\,\Omega_{\gamma,0}\,\rs(k) \,.
\end{equation}
% The amplitude of $F_{\rm rad}$ peaks on  scales just inside the
% Hubble radius and becomes scale-invariant on smaller scales.
%
where, for modes which are well inside the horizon at the end of the
radiation-dominated era ($k\eta_{eq}\gg1$), we have
\cite{Ananda:2006af}
\begin{equation}
\label{frad} F_{\rm rad} = \frac{8}{3} \left( \frac{216^{2}}{\pi^{3}}
\right)
 8.3 \times 10^{-3} f_{ns}
\end{equation}
and $f_{ns}$ is weakly-dependent on the spectral tilt.
$f_{ns}\approx1$ if the density perturbation is scale-invariant
\cite{Ananda:2006af}, but becomes slightly smaller than one for a
red spectrum (e.g, $f_{ns} \approx 0.97$ for $n_s=0.9$) and bigger
than one for a blue spectrum (e.g, $f_{ns} \approx 1.05$ for
$n_s=1.1$).

% v5
It is also possible to consider the spectrum of gravitational waves generated by density perturbations with a sharply peaked power spectrum \cite{Ananda:2006af,Saito:2008jc}. Considering a delta-function power spectrum, $P(k)=(4/9)\Delta_{\cal R}^2(k_p)\delta(\ln(k/k_p))$, the resulting gravitational wave spectrum is described by a sharply rising spectrum for $k<k_p$ \cite{Saito:2008jc}
\begin{equation}
 \label{OmegaGWdelta}
 \Omega_{gw,0}(k)= 29 \,\Omega_{\gamma,0}\,\rs(k_p) \left( \frac{k}{k_p} \right)^2 \,,
\end{equation}
with an abrupt cut-off for $k>2k_p$.

In the following numerical estimates we take $F_{\rm
rad}\approx30$
% v5
in Eq.~(\ref{hr1})
corresponding to an approximately scale-invariant
spectrum of scalar perturbations, $n_s\approx1$. This is expected
to be a conservative lower bound on $F_{\rm rad}$ for the blue
spectra with $n_s>1$ required to produce a detectable background
of gravitational waves on scales much smaller than the CMB scale.
In the rest of this section we show, how Eq.~(\ref{hr1}) enables us
to use constraints on the stochastic background of gravitational
waves generated during the radiation era, $\Omega_{gw,0}(k)$, to place upper bounds on
the primordial density perturbation on the corresponding scales, $\Delta_{\cal R}^2(k)$.
% v5
In addition Eq.~(\ref{OmegaGWdelta}) indicates how observational constraints on $\Omega_{gw,0}(k)$ at a given wavenumber $k$ also places a weaker bound on the primordial density perturbation, $\Delta_{\cal R}^2(k_p)\propto (k_p/k)$, at higher wavenumbers, $k_p>k$.
Our results are presented graphically in Figure~\label{fig:radiation}.

\begin{figure}
\scalebox{0.2}{\includegraphics{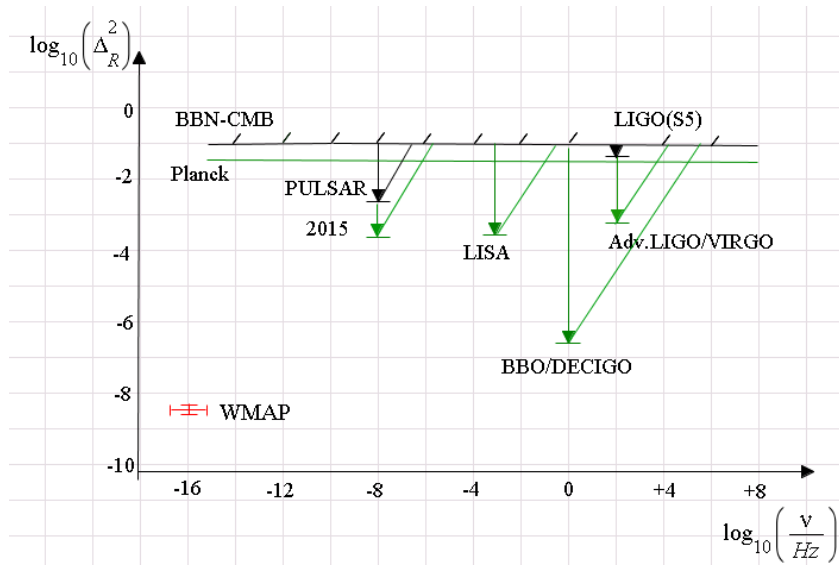}}
\caption{Constraints on the primordial density perturbation, $\Delta_{\cal R}^2$, obtained from
gravitational waves produced during the radiation era, using Eq.~(\ref{hr1}). Black lines
denote current constraints from gravitational waves detectors and
BBN. Green lines denote constraints expected from future
gravitational waves detectors.
% v5
Diagonal lines show the bounds on $\Delta_{\cal R}^2$ obtained using Eq.~(\ref{OmegaGWdelta}) for wavenumbers $k_p>k$.
WMAP gives a direct measurement (shown in red) of the primordial density perturbation on very low frequencies.
}
\label{fig:radiation}
\end{figure}

\subsection{Cosmological density constraints}

\subsubsection{Constraint from BBN}
\label{sectBBN}

If the energy density carried by gravitational waves at the time of
primordial big bang nucleosynthesis (BBN) were large, the abundances
of the light nuclei produced would be altered with respect to the
predictions of standard BBN. Hence, BBN can be used to constrain the
total energy carried by gravitational waves at the time of
nucleosynthesis ($T\simeq 1$~MeV) \cite{Cyburt:2004yc}.

Primordial abundances of the light elements, usually quoted as a
bound on the effective number of relativistic species at the time of
BBN, gives the 95\% c.l.~upper bound on a primordial gravitational
wave background \cite{Abbott:2006zx}
%
%photon must be the dominant energy density in the early universe
%to become consistent with present baryon-to photon ratio. Hence
%BBN places an upper limit on induced gravitational waves:
%begin{equation}
%\label{bbnc1} \Omega_{gw,0}<0.1\,\Omega_{\gamma,0}
%\end{equation}
\begin{equation}
 \label{bbnc1} \Omega_{gw,0}< 1.5\times 10^{-5}
\end{equation}
Substituting this bound into Eq.~(\ref{hr1}) gives
\begin{equation}
\label{bbnc2} \rsn < 0.1 \left(\frac{F_{rad}}{30}\right)^{-\frac{1}{2}}
\,.
\end{equation}
This denotes the upper bound on the primordial density perturbation
on the Hubble scale at the time when the gravitational waves are
generated.

Although Eq.~(\ref{bbnc2}) is only a weak limit on the primordial
density perturbation, it is applies across a wide range of length
scales. It applies on scales which are smaller than the Hubble
scale at the epoch of BBN, which from Eq.~(\ref{fT}) corresponds
to frequencies $\nu>10^{-10}$ Hz today, and scales which are
larger than the Hubble scale at the start of the
radiation-dominated era. This is model-dependent, but in an
inflationary cosmology this would be the Hubble scale at the end
of reheating after inflation, which could be as large as $\nu\sim
10^8$~Hz for $T\sim 10^{16}$~GeV. If inflation occurs at lower
energy scales the reheating temperature could be much lower.

The relationship between the primordial density perturbation on the
CMB scale, $\rsn(k_{CMB})$, and on an arbitrary scale, $\rsn(k)$,
can be written as
\begin{equation}
\label{ns1k}
\rsn(k)=\rsn(k_{CMB})\times\left(\frac{k}{k_{CMB}}\right)^{\overline{n}_s-1}
\end{equation}
where $\overline{n}_s$ describes the effective tilt between the
scale $k$ and CMB scales, where $\rsn$ is directly observed for
$\rsn(k_{CMB})=2.5\times10^{-9}$ for $k_{CMB}=0.002$~Mpc$^{-1}$
\cite{Komatsu:2008hk}.
In terms of frequency we have
% \begin{equation}
% \label{ns} \rsn=\rsn
% (k_{CMB})\times\left(\frac{\nu}{\nu_{CMB}}\right)^{n_{s}-1}
% \end{equation}
% (\ref{ns}) gives $n_{s}$ in terms of  $\rsn$ for different scales:
\begin{equation}
\label{ns1} \overline{n}_s =
1+\frac{\log\left(\frac{\rsn}{\left(\rsn(k_{CMB})\right)}\right)}{\log\left(\frac{\nu}{\nu_{CMB}}\right)}
 \,,
\end{equation}
where $\nu_{CMB}\approx10^{-18}$~Hz. Therefore (\ref{bbnc2}) can be
interpreted as a constraint on the value of the effective spectral
index, $\overline{n}_s$, across a wide range of scales.

According to WMAP5 data \cite{Komatsu:2008hk}, the spectrum of
primordial density perturbations is red ($n_{s}$ less than unity)
on CMB scales but, except for the constraints from primordial
black holes (PBHs), we have no restrictions on the value of
$\overline{n}_s$ on scales much smaller than 1~Mpc. We obtain the
tightest constraint on $\overline{n}_s$ from a bound such as
Eq.~(\ref{bbnc2}) applied to the smallest possible wavelength.
%
% This mode depends on the energy scale after reheating after
% inflation and its highest possible value is $10^{16}$ GeV
% \cite{Kolb:1990vq}. Equation (\ref{fT}) shows that the energy scale
% $T\sim10^{16}$ GeV corresponds to $10^{8}$ Hz.
%
For $\nu\simeq 10^{8}$~Hz from (\ref{bbnc2}) and (\ref{ns1}) we find
% v4

\begin{equation}
\label{nsrbbn} \overline{n}_s < 1.29-\frac{1}{52}\log_{10}
\left(\frac{F_{rad}}{30}\right)\,.
\end{equation}
Equation (\ref{nsrbbn}) corresponds to the maximum possible reheat
temperature after inflation. However if we consider a smaller
reheat temperature, for instance $T\simeq 10^{8}$~GeV
corresponding to $\nu\simeq 1$~Hz, we get
$\overline{n}_s<1.44-\frac{1}{36}\log_{10}
\left(\frac{F_{rad}}{30}\right)$.

% v4 new subsection
\subsubsection{Constraint from CMB}
\label{sectCMB}

A very similar bound on the effective energy density of primordial gravitational waves can be obtained around the time of last scattering of the cosmic microwave background. Again a limit on the number of massless neutrino species~\cite{Pierpaoli:2003kw} can be translated into a bound on the gravitational wave background~\cite{Smith:2006nka}. Unlike the BBN constraint, however, the CMB constraint depends upon the nature of inhomogeneous perturbations about the average density. For a gravitational wave background produced from a Gaussian random field of primordial density perturbations on small scales, we expect the effective energy density on long wavelengths (on scales of order $100$~Mpc) to be independent of the density perturbations on this scale. Thus long wavelength perturbations of the gravitational wave background are non-adiabatic and Smith {\em et al} \cite{Smith:2006nka} give a 95\% c.l.~bound
\begin{equation}
\Omega_{gw,0} < 1.3 \times 10^{-5}
%\left( \frac{72 {\rm km\ s}^{-1}\ {\rm Mpc}^{-1}}{H_0} \right)^2
 \,,
\end{equation}
for a ``homogeneous'' gravitational wave background. This is marginally stronger than the BBN constraint (\ref{bbnc1}). It gives effectively the same constraint on the primordial density perturbation (\ref{bbnc2}), and the effective spectral index (\ref{nsrbbn}), but extends to longer wavelengths $\sim 10^{-15}$~Hz, corresponding to scales inside the Hubble scale at the time of last-scattering.

Future data from CMB experiments such as Planck and the proposed CMBPol mission are expected to improve the CMB bound. For Planck the expected bound corresponds to $\Omega_{gw,0} < 2.7 \times 10^{-6}$ \cite{Smith:2006nka} which would bound
\begin{equation}
\rsn < 0.04 \left(\frac{F_{rad}}{30}\right)^{-\frac{1}{2}} \,.
\end{equation}

\subsection{Constraints from ground-based detectors}

\subsubsection{Current LIGO/VIRGO}

We can obtain a tighter constraint on the primordial density
perturbation on scales probed by direct detectors, such as the Laser
Interferometer Gravitational Wave Observatory (LIGO) \cite{Abbott:2006zx} and
gravitational wave detector at the European gravitational
observatory (VIRGO) \cite{Acernese:2008zzf}.
LIGO's maximum sensitivity is around a frequency, $\nu=100$~Hz.
% v4
The latest results from the LIGO S5 science run give a bound
% \cite{Abbott:2006zx,Acernese:2008zzf},
on the energy density of
gravitational waves on this scale \cite{LIGOS5}
\begin{equation}
\Omega_{gw,0} < 6.9 \times 10^{-6} \,.
\end{equation}
Hence from Eq.~(\ref{hr1}) the constraint
on the density perturbation on the LIGO/VIRGO scale is
\begin{equation}
\label{ligoc1} \rsn <
0.07\left(\frac{F_{rad}}{30}\right)^{-\frac{1}{2}} \,.
\end{equation}
This is a slightly tighter bound than that currently obtained from BBN and the CMB,
Eq.~(\ref{bbnc2}), however unlike the BBN and CMB bound it only applies to LIGO/VIRGO
scales. The corresponding constraint on $\overline{n}_s$ on this
scale comes from Eq.~(\ref{ns1})
% V4

\begin{equation}
\label{nsL} \overline{n}_s < 1.37 -\frac{1}{40}\log_{10}
\left(\frac{F_{rad}}{30}\right)\,.
\end{equation}
%Because gravitational waves have not been detected with LIGO, on
%this scale we have the above constraint on the primordial density
%perturbation and $n_{s}$.

% So far we found the constraints for the primordial density
% perturbation with current detectors. For the rest of this section
% we will determine, the same constraints with future gravitational
% waves detectors. It will be shown how constraints on the
% primordial density perturbation become tighter, if a stochastic
% background of  gravitational waves generated during the radiation
% era are not observed by future detectors.

\subsubsection{Advanced LIGO/VIRGO}

Advanced LIGO/VIRGO will give us an improved constraint on a
stochastic background of gravitational waves on the same scales
\cite{Sigg:2008zz,Acernese:2008zzf}. The smallest density of
gravitational waves which could be detected by Advanced LIGO/VIRGO
is $10^{3}$ times smaller than current LIGO/VIRGO bounds.
Considering the smallest detectable energy density
$\Omega_{gw,0}<10^{-9}$ in Eq.~(\ref{hr1}) returns
% v4
\begin{equation}
\label{aligoc1} \rsn <
8\times10^{-4}\left(\frac{F_{rad}}{30}\right)^{-\frac{1}{2}}
\end{equation}
Equation (\ref{ns1}) gives the expected constraint on
$\overline{n}_s$ from Advanced LIGO/VIRGO (taking $\nu=100$ Hz)
% v4
\begin{equation}
\label{nsaL}
 \overline{n}_s < 1.27 -\frac{1}{40}\log_{10}
\left(\frac{F_{rad}}{30}\right)\,.
\end{equation}

\subsection{Constraints from LISA}

The Laser Interferometer Space Antenna (LISA) is the first
gravitational wave detector planned in space and is the most
sensitive detector currently planned.
% v4 DW 16/10/09
Assuming LISA's instrumental noise is well-behaved~\cite{Hogan:2001jn}, it could detect a stochastic background of gravitational waves at a level
$\Omega_{gw,0}\sim10^{-11}$ at frequencies $\nu_{LISA}\sim10^{-3}$ Hz \cite{Hughes:2007xm,Sathyaprakash:2009xs}. However the sensitivity of LISA leads to many potential overlapping sources and hence the problem of source confusion. In particular the astrophysical background from unresolved extra-galactic white-dwarf binaries is expected to limit LISA's ability to distinguish a primordial gravitational wave background to~\cite{Hogan:2001jn,Cornish:2001bb}
\begin{equation}
 \Omega_{gw,0} < 10^{-10} \,.
\end{equation}
The corresponding upper bound on the primordial density
perturbation on LISA scales comes from (\ref{hr1}):
\begin{equation}
\label{lisac1}
\rsn<3\times10^{-4}\left(\frac{F_{rad}}{30}\right)^{-\frac{1}{2}}
\end{equation}
The  constraint on $\overline{n}_s$ on LISA scales comes from
Eq.~(\ref{ns1})
% v4

\begin{equation}
\label{nsLI} \overline{n}_s < 1.34 -\frac{1}{30}\log_{10}
\left(\frac{F_{rad}}{30}\right)\,.
\end{equation}
This is a slightly  weaker bound on the effective spectral index
compared with Advanced LIGO, as LISA is sensitive on length scales
much larger than LIGO scales.

\subsection{Constraints from BBO/DECIGO}

The Big Bang Observer (BBO) \cite{Corbin:2005ny} and the DECi-hertz
Interferometer Gravitational wave Observatory (DECIGO)
\cite{Kawamura:2008zz} are ambitious proposals for future
space-based observatories to detect cosmological gravitational
waves. They should be able to detect a stochastic background of
gravitational waves down to an effective energy density
$\Omega_{gw,0}\approx10^{-16}$ at $\nu_{BBO}\approx1$ Hz.
% v5
This waveband is chosen to avoid the confusion noise due to white dwarf binary mergers which cuts off above $0.2$~Hz. The designs of BBO and DECIGO are based on the requirement to identify and remove the remaining foregrounds from neutron star and black hole binaries \cite{Cutler:2005qq}.

If induced
gravitational waves during the radiation era are not detected with
BBO/DECIGO, then we will be able to place a tight constraint on the
primordial density perturbation and hence $\overline{n}_{s}$ on this
scale (1 Hz). From (\ref{hr1}) and (\ref{ns1}) we obtain
\begin{equation}
\label{BBOc1}
 \rsn < 3\times10^{-7}\left(\frac{F_{rad}}{30}\right)^{-\frac{1}{2}} \,,
\end{equation}
and hence
% v4

\begin{equation}
\label{nsBBO} \overline{n}_s < 1.11-\frac{1}{36}\log_{10}
\left(\frac{F_{rad}}{30}\right) \,.
\end{equation}

\subsection{Constraints from pulsar timings}

Analysis of  pulse data from pulsars shows that they are very stable
clocks. Measurement of timing residuals, which is the difference
between the observed time of arrival and predicted time of arrival,
can in principle be used to directly detect gravitational waves
passing between the pulsar and the observer
\cite{Jenet:2005pv,Jenet:2006sv}. The data from current observations
of an array of pulsars places an upper bound on the stochastic
background of gravitational waves, with periods comparable to the
total observation time span. This is typically 1-10 years, and
corresponds to $10^{-8}$-$10^{-9}$ Hz. For $\nu=10^{-8}$ Hz, the
constraint on the present density of gravitational waves is
\cite{Jenet:2006sv}
\begin{equation}
 \label{pulsar1} \Omega_{gw,0} < 4\times10^{-8} \,.
\end{equation}
Substituting (\ref{pulsar1}) in (\ref{hr1}), gives us the current
constraints on the primordial density perturbation
\begin{equation}
\label{pulsardr} \rsn <
5 \times 10^{-3} \left(\frac{F_{rad}}{30}\right)^{-\frac{1}{2}} \,.
\end{equation}
The constraint on $\overline{n}_s$ comes from (\ref{hr1})
% v4

\begin{equation}
\label{pulsarnr} \overline{n}_s < 1.63 -\frac{1}{20}\log_{10}
\left(\frac{F_{rad}}{30}\right)\,.
\end{equation}

Saito and Yokoyama \cite{Saito:2008jc} have recently used similar
constraints, on the induced gravitational wave background from
pulsar timing arrays, to rule out the large amplitude of primordial
density perturbations required to produce any significant number of
primordial black holes in the intermediate mass range,
$4\times10^{2}M_\odot\leq M_{PBH} \leq 5\times 10^3M_\odot$,
corresponding to $8\times10^{35}\ {\rm g}\leq M_{PBH} \leq 10^{37}
{\rm g}$, which from Eq.~(\ref{PBHMT}) would have formed at
temperatures $T \approx 3-10\ {\rm MeV}$.

Future pulsar timing will give a better constraint. If gravitational
waves are not detected, the upper limit, based on timing 20 pulsars
over 5 years, would be $\Omega_{gw,0}<10^{-10}$ \cite{Jenet:2006sv}.
{}From (\ref{hr1}), the future constraint on the
primordial density perturbation in five years time for $\nu=10^{-8}$
Hz would be
\begin{equation}
\label{pulsardrf} \rsn <
3\times10^{-4}\left(\frac{F_{rad}}{30}\right)^{-\frac{1}{2}} \,.
\end{equation}
The future constraint on $\overline{n}_s$ would then be
% v4

\begin{equation}
\label{pulsarnrfuture} \overline{n}_s < 1.50
-\frac{1}{20}\log_{10} \left(\frac{F_{rad}}{30}\right)\,.
\end{equation}

\section{Conclusion}

Despite remarkable recent progress in astronomical observations
mapping density perturbations on large scales (10-1000~Mpc) in our
Universe, we know little about the primordial distribution of matter
on much smaller scales. This is due to Silk damping and
free-streaming of relativistic particles in the early universe, and
subsequent non-linear evolution of matter perturbations at much
later times. The only constraints on these scales come from
gravitational relics of the very early universe. Previous work has
focussed on the possible formation of primordial black holes from
large over-densities.

In this paper we have shown how limits on a stochastic background of
gravitational waves can be used to place limits on density
perturbations in the early radiation-dominated era of the standard
hot big bang cosmology.

BBN and CMB limits on a primordial gravitational wave background places
only a weak constraint on the amplitude of primordial density
perturbations, $\rsn<0.1$, but this applies across a wide range of
frequencies, from $10^{-15}$~Hz to frequencies as high as
$10^8$~Hz, depending on the maximum temperature at the start of
the radiation-dominated era. By contrast, gravitational wave
detectors such as LIGO and VIRGO place slightly tighter bounds, currently
$\rsn<0.07$, but only over a narrower range determined by the
frequency response of the detector.

Future gravitational wave experiments offer the prospect of much
tighter bounds on, or a detection of, a stochastic gravitational
wave background and hence the primordial density perturbation on
small scales. A space-based experiment such as LISA could detect
gravitational waves produced by density perturbations $\rsn\sim
10^{-4}$, and future data from pulsar timing arrays could have
similar sensitivity. The most ambitious current proposed
gravitational wave observatories including BBO and DECIGO offer the
prospect of detecting gravitational waves as small as
$\rsn\sim10^{-7}$.

If gravitational wave background is not detected by these
experiments it would imply that the primordial power spectrum
remains close to scale-invariant, or decreases in power on small
scales, $ \overline{n}_s <1.29$, which provides a valuable new
constraint on models for the origin of structure. Nonetheless it
remains a challenge to design an experiment that could detect
gravitational waves produced by primordial density perturbations
of the same power, $\rsn\sim10^{-9}$, as seen on the largest
scales in the universe today.

We have assumed the simplest expansion history of the universe,
being radiation dominated from very early times, corresponding to
temperatures as high as $10^{16}$~GeV. If the early universe has a
more complicated history, the constraints may be altered. A period
of inflation is expected to dilute pre-existing gravitational waves
on sub-Hubble scales, without generating a significant additional
background~\cite{Arroja:2009sh}. On the other hand although an early
matter-dominated era before BBN, such as the reheating or preheating
after inflation, would also dilute any gravitational waves that had
already been generated, it could itself produce significant tensor
metric perturbations on scales that re-enter the Hubble scale during
an early matter-dominated era \cite{Assadullahi:2009nf}. We leave a
more detailed investigation of constraints on the primordial density
perturbation in more general cosmological scenarios to future work.

\section*{Acknowledgements}

We are grateful to Marco Bruni, Karen Masters and Misao Sasaki for
useful comments. HA is supported by the British Council (ORSAS). DW
is supported by the STFC.

%\bibliographystyle{utphys}
%\bibliography{mybib2}

\end{document}